\documentstyle[preprint,prd,aps,epsfig]{revtex}

\tighten
\begin{document}

\newcommand{\gsim}{\gtrsim}
\newcommand{\lsim}{\lesssim}
\newcommand{\psim}{\mbox{\raisebox{-1.0ex}{$~\stackrel{\textstyle \propto}
{\textstyle \sim}~$ }}}
\newcommand{\vect}[1]{\mbox{\boldmath${#1}$}}
\newcommand{\lmk}{\left(}
\newcommand{\rmk}{\right)}
\newcommand{\lnk}{\left\{ }
\newcommand{\nn}{\nonumber}
\newcommand{\rnk}{\right\} }
\newcommand{\lkk}{\left[}
\newcommand{\rkk}{\right]}
\newcommand{\lla}{\left\langle}
\newcommand{\p}{\partial}
\newcommand{\rra}{\right\rangle}
\newcommand{\lop}{\left\|}
\newcommand{\rop}{\right\|}
\newcommand{\vex}{{\vect x}}
\newcommand{\veR}{{\vect R}}
\newcommand{\uve}{{\hat{\vect e}}}
\newcommand{\veg}{{\vect\gamma}}
\newcommand{\beq}{\begin{equation}}
\newcommand{\eeq}{\end{equation}}
\newcommand{\beqa}{\begin{eqnarray}}
\newcommand{\eeqa}{\end{eqnarray}}
\newcommand{\lab}{\label}

\draft
\title{Polarization Signal of
Distant Clusters and Reconstruction of Primordial Potential Fluctuations }
\author{Naoki Seto and Misao Sasaki}
\address{Department of Earth and Space Science,
Osaka University,
Toyonaka 560-0043
}

\date{\today}
\maketitle
\begin{abstract}
We examine the  polarization signal of the cosmic microwave background
radiation   associated with
 distant  clusters.
 The polarization is  induced by the Thomson scattering of microwave
 photons with ionized gas of clusters and
contains information of quadrupole temperature anisotropies
 observed at the clusters. The three-dimensional map of the signal are
 expressed  in terms of  the spin-weighted harmonics for its angular
 dependence.  Its radial dependence is expanded perturbatively with
 respect to the distances (equivalently redshifts) to the clusters.
 The  independent 
 information that we can extract out from the map is clarified
 explicitly.
 
\end{abstract}

\section{Introduction}
Reconstruction of large-scale density fluctuations  is an important
topic in modern  cosmology.   Nowadays there are mainly two methods for
this 
reconstruction.  The redshift surveys of galaxies  probe the density
(more strictly the number density of galaxies) fluctuations  around our
local (nearby)
universe. The anisotropies of the
cosmic microwave background (CMB) contain 
information of
potential fluctuations at the last scattering surface (LSS).  The 
redshift survey has been extended to 
 deeper and deeper universe and has brought us enormous information
of the three-dimensional matter distribution. On the other hand,
the CMB data, although carrying information of the deepest universe,
are of two-dimensional nature as they are.
It would be therefore very interesting to explore the possibilities
of extracting out the three-dimensional information of
primordial fluctuations around the LSS.

The linear polarization of CMB is induced at a cluster of galaxies by
the Thomson scattering of CMB photons with the ionized gas (electron)
of the cluster \cite{zelsun,koso}.  The polarization signal is related
to one component of the CMB quadrupole moment observed at the cluster.
Using this fact, Sazonov \& Sunyaev \cite{ss} predicted the linear
polarization signal of nearby clusters 
with approximations that  the  temperature
anisotropies observed at these clusters are the same as that observed at
our galaxy.     
They used quadrupole moment of CMB measured by the Cosmic Background
Explorer satellite \cite{kogu}.  

As pointed out by Kamionkowski \& Loeb \cite{kl}, the LSS
 of an observer (cluster)
  depends on the position of the observer. Therefore the LSS of a distant
cluster 
is shifted from our LSS \footnote{The LSS of a distant cluster is also
smaller than ours due to the light-cone effect.} and its polarization signal  would
probe three dimensional information of potential fluctuations around our
LSS.  Kamionkowski \& Loeb noticed this effect and commented
that the polarization signal of clusters might be used to  reduce
cosmic variance limitation of large-scale power spectrum. 

Considering further the shift of LSS mentioned above,  we can
expect that correction terms in
the polarization signal proportional to the redshift of 
clusters  $\propto z$ would reflect the first radial derivative of the
potential fluctuations at our LSS. In the same manner the
 signal proportional to the square of
redshift $\propto z^2$ would reflect the second radial derivative
and so on.
In this paper we calculate the three-dimensional map of
polarization signal induced by the quadrupole
temperature anisotropies  of CMB at distant clusters (see also
Ref.\cite{zal1,kami1,hu1}). Then we clarify the
information we can extract out from the map.  There are
many astrophysical or cosmological effects on the polarization
of the CMB, such as,  the peculiar velocities of clusters or
reionization of the universe (see {\it e.g.} Ref \cite{ss,kl}). We
do not go into these issues but investigate a basic problem
about reconstruction of the primordial potential
fluctuations from the quadrupole anisotropies of CMB observed
at distant clusters.

\section{Formulation}
\subsection{Polarization Induced by the  Thomson Scattering}
Polarization of CMB in the direction of a cluster reflects the 
 temperature
anisotropy of CMB observed at the cluster. In this subsection we briefly
review this effect following Kosowsky \cite{koso}.
Let us consider a nearly monochromatic plane electromagnetic wave
 propagating to the $z$-direction. We denote its electric field as  
\beq
E_x=a_x(t) \cos[\omega_0t-\theta_x(t)],~~~
E_y=a_y(t) \cos[\omega_0t-\theta_y(t)].
\eeq
Here the amplitudes $a_x(t)$,  $a_y(t)$ and the phases $\theta_x(t)$,
$\theta_x(t)$ are nearly 
constant in the oscillation time scale $\omega_0^{-1}$ of the wave.  
The Stokes parameters  characterize the polarization of radiation
field and are defined as follows (Chandrasekhar \cite{chandra}, Rybicki
 \& Lightman \cite{radi})
\beqa
I&=&\lla a_x^2\rra+\lla a_y^2\rra,\\
Q&=&\lla a_x^2\rra-\lla a_y^2\rra,\\
U&=&\lla 2 a_x a_y \cos(\theta_x-\theta_y)\rra,\\
V&=&\lla 2 a_x a_y \sin(\theta_x-\theta_y),\rra,
\eeqa
where brackets $\lla\cdot\rra$ represent time averages.

Next we calculate the polarization induced by the Thomson scattering.
The cross section of  this process is determined by the
polarization vector of the incident wave $\hat{\epsilon}'$ and
that of the scattered wave $\hat{\epsilon}$ as \cite{chandra}
\beq
\frac{d\sigma}{d\Omega}
=\frac{3\sigma_T}{8\pi}|\hat{\epsilon}\cdot\hat{\epsilon}'|^2,\lab{thom}
\eeq
where $\sigma_T$ is the Thomson cross section (figure 1).
When the incident wave is natural ($Q=U=0$), it is completely
described 
by the angular dependence of the intensity $I'(\theta,\phi)$. 
Using the above 
cross section (\ref{thom}) and  transformation of the Stokes parameters
under rotation of coordinate systems, we obtain the Stokes 
 parameters $I$,  $Q$
and $U$ for the scattered wave as
\beqa
I&=&\frac{3\sigma_T}{16\pi}\int d\Omega(1+\cos^2\theta) I'(\theta,\phi),\\
Q&=&\frac{3\sigma_T}{16\pi}\int 
d\Omega\sin^2\theta \cos(2\phi) I'(\theta,\phi),\lab{pq}\\
U&=&-\frac{3\sigma_T}{16\pi}\int d\Omega\sin^2\theta \sin(2\phi)
I'(\theta,\phi)\lab{pu}.
\eeqa
The parameter $V$ which characterizes the circular polarization remains
zero in the  Thomson scattering.

Let us consider a cluster of galaxies (as a cloud of ionized gas)  whose
optical depth $\tau$ for the Thomson 
scattering is much smaller than unity.   In this optically thin limit
we  obtain the total Stokes 
parameters $Q$ and $U$ that are induced by the  cluster by   replacing
$\sigma_T$ with 
the optical depth $\tau$ in  equations (\ref{pq}) and (\ref{pu}).
To evaluate the Stokes parameters we expand the angular
        dependence of the intensity $I'$
        in terms of the spherical harmonics as follows
\beq
I'(\theta,\phi)
             =\sum_{lm} {\cal I'}_{lm} Y_{lm}(\theta,\phi).
\eeq
Then equations (\ref{pq}) and (\ref{pu}) become
\beq
Q=\frac{3\tau}{4\pi}\sqrt{\frac{2\pi}{15}}{\rm Re}({\cal I'}_{22}),~~~
U=-\frac{3\tau}{4\pi}\sqrt{\frac{2\pi}{15}}{\rm Im}({\cal I'}_{22}),
\eeq
or these are simply  combined as 
\beq
Q-iU=\frac{3\tau}{4\pi}\sqrt{\frac{2\pi}{15}}{\cal I'}_{22}.
\eeq
Therefore, by measuring the polarization parameters $(Q,U)$  
in the direction of a
cluster,  we can, in principle, measure the 
quadrupole anisotropy ${\cal I'}_{22}$ of the CMB observed at the cluster.
In reality the incident waves have some degree of polarization. The
primordial contribution at small $l$ is expected to be much smaller than
the temperature anisotropy for typical CDM models, but the total
magnitude can become nonnegligible depending 
on the reionization history ({\it e.g.} Ref.\cite{zal1,hu1,zal2}). 
Furthermore, one cannot deny the possibility of a large 
tensor contribution to the CMB quadrupole.
In this paper, however, we ignore such contributions and 
focus on a somewhat fundamental problem; what
information of the primordial potential field
can we reconstruct by using the quadrupole moment ${\cal T}_{22}$
 observed at different places? 

\subsection{Temperature Anisotropies of the CMB}
In an optically thin universe after decoupling,  propagation of the
gauge invariant brightness (temperature) perturbation
${\cal{T}}_s$ on the Newtonian hypersurface  is written as (Kodama \&
Sasaki \cite{kodsas})
\beq
\frac{d}{d\eta}[{\cal{T}}_s(\eta,\vex(\eta),\veg)+\Psi(\eta,\vex(\eta))]
=\frac{\p}{\p \eta}(\Psi-\Phi),
\eeq
where $\Psi$ and  $\Phi$ are the  Newtonian and spatial
          curvature perturbation, and $\vex(\eta)$ represents the
          null geodesic of the propagating photon that is 
          parameterized by the conformal time $\eta$. The temperature
          perturbation ${\cal{T}}_s$  is related to the  intensity
          perturbation $\Delta I $ as ${\cal{T}}_s=\Delta I/(4I)$.

This equation can be formally integrated and the function ${\cal{T}}_s$
at an epoch $\eta$ is written in terms of quantities at decoupling
 $\eta_{dec}$ as
\beqa
{\cal{T}}_s(\eta,\vex(\eta),\veg)&=&{\cal{T}}_s(\eta_{dec},\vex(\eta_{dec}),\veg)+\Psi(\eta_{dec},\vex(\eta_{dec}))-\Psi(\eta,\vex(\eta))\nn\\
& &+2\int_{\eta_{dec}}^{\eta}\frac{\p}{\p
\eta'}\Psi(\eta',\vex(\eta'))d\eta'.
\eeqa
As the anisotropic pressure perturbation is negligible after decoupling,
we have put $\Phi=-\Psi$.

For adiabatic perturbations the large scale
behavior of  the brightness perturbation  $\cal{T}$  at decoupling is
approximately given by the potential field at the same time
(Ref.\cite{sw}) 
\beq
{\cal{T}}_s(\eta_{dec},\vex(\eta_{dec}),\veg)
\simeq-\frac23 \Psi (\eta_{dec},\vex(\eta_{dec})).
\eeq
This is an efficient approximation to discuss small-$l$ (large
angle) temperature anisotropies ({\it e.g.}, ref.\cite{gouda}).
Thus we consider two effects for anisotropies  ${\cal T}_s$ seen at
clusters that are 
known as (i) the Sachs-Wolfe (SW) effect ${\cal{T}}_{SW}$ and (ii) the
Integrated-Sachs-Wolfe (ISW) effect  ${\cal{T}}_{ISW}$,
\beqa
{\cal{T}}_{s}&=&{\cal{T}}_{SW}+{\cal{T}}_{ISW},\\
{\cal{T}}_{SW}(\eta,\vex(\eta),\veg)&\equiv& \frac13 \Psi
(\eta_{dec},\vex(\eta_{dec})),\\
{\cal{T}}_{ISW}(\eta,\vex(\eta),\veg)&\equiv
&2\int_{\eta_{dec}}^{\eta}\frac{\p}{\p\eta'}\Psi(\eta',\vex(\eta'))d\eta'.
\eeqa
Note that the anisotropy ${\cal{T}}_{s}(\eta,\vex,\veg)$ is 
defined on the shear-free (Newtonian) hypersurface. Nevertheless, 
since the coordinate gauge transformation affects only the monopole and
dipole components of the anisotropy, our analysis does not depend on 
the choice of the hypersurface.

\subsection{Transformation of Coordinate Systems}
In this section we discuss the relation between two spherical coordinate
systems centered at different places.  We consider a
 homogeneous and isotropic universe whose  metric  is
given by
\beq
ds^2=a(\eta)^2(-d\eta^2+dr^2+ f(r)^2
(d\theta^2+\sin^2\theta d\phi^2))\lab{metric},
\eeq
where the function $f(r)$ depends on the spatial curvature radius $B$
and is
defined as
\[ f(r)=\cases{
B\sin\lkk\frac{r}{B}\rkk & Closed~model, \cr
r & Flat~model, \cr
B\sinh\lkk\frac{r}{B}\rkk & Open~model. \cr
}\]
A null geodesic from (or into) the origin $r=0$ along
$\theta=$const. and $\phi=$const. is trivially solved in 
this  coordinate system  and calculation of the temperature anisotropy
 ${\cal T}(O',\veg)$
observed at a cluster $O'$ is very easy using  a coordinate
system centered 
at $O'$ (hereafter call the $O'$-system). However, the linear  potential
field $\Psi$ is more informative  
when it is expressed in terms of the coordinate system centered at the
Earth $O$ (the $O$-system). Therefore,  we need the
relation between the  spherical coordinate systems centered at different
points $O$ and $O'$.
 We denote the position of
the cluster $\overrightarrow{OO'}\equiv d\uve_c$ ($d$: distance to the
cluster).

We first set the direction of $\theta=0$  parallel to the
direction of the cluster, $\overrightarrow{OO'}$,  both for the two
systems. Then,
due to the rotational symmetry around the axis $\overrightarrow {OO'}$,
we can take
the same  azimuthal angle $\phi$  in the two 
systems.\footnote{The direction of $\phi=0$ is arbitrary. See also
\S2.E.} In the following  
we relate the radial distance $R$ and the  angle $\Theta$ in the
$O$-system with the corresponding $r$ and $\theta$ in the  $O'$-system. 
We examine
the case of a closed universe, but the results can be straightforwardly
extended to flat and open models. 

Since the two-dimensional space spanned  by $(r,\theta)$ with the metric
 (\ref{metric}) can be embedded in the three-dimensional Euclidean space
$(X_1,X_2,X_3)$ as a two-sphere of radius $B$, 
the correspondence of the two coordinate systems $(R,\Theta)$ and
 $(r,\theta)$ is 
simply obtained by rotation of the $O'$-system around the $X_2$-axis with
the angle $\alpha\equiv d/B$ (see figure 2). Then we obtain the following
embedding relation for the three-dimensional 
coordinates $(X_1,X_2,X_3)$ as
\beqa
{X_1\over B}&=&\sin \frac{R}B\cos\Theta=\sin \frac{r}B \cos\theta
\cos\alpha+\cos\frac{r}B
\sin\alpha,\\
{X_2\over B}&=&\sin\frac{R}B \sin\Theta=\sin\frac{r}B\sin \theta,\\
{X_3\over B}&=&\cos \frac{R}B
=\cos \frac{r}B \cos\alpha-\sin\frac{r}B \cos\theta
\sin\alpha.
\eeqa
Using the above, an event $(\eta,R,\Theta,\phi)$ in the $O$-system 
can be perturbatively expressed in terms of the corresponding event
 $(\eta,r,\theta,\phi)$ in the $O'$-system as
\beqa
\eta&=&\eta,
\nonumber\\
R&=&r+d\cos\theta+O(d^2),
\nonumber\\
\Theta&=& \theta -\frac{d}{B}\sin\theta\cot\frac{r}{B}+O(d^2),
\nonumber\\
\phi&=&\phi.
\label{OOprel}
\eeqa
Here we have assumed $r\gg d$. This assumption formally breaks down
when we calculate the ISW contribution near the cluster (see the next
subsection). Nevertheless, this contribution is found to be negligible
at the $d^1$-st order for  $\eta_0-\eta_{dec}\gg d$,
 where $\eta_0$ is the conformal time at present. 

Using the above equations (\ref{OOprel}),
the potential perturbation $\Psi_{O'}(\eta,r,\theta,\phi; \uve_c)$ in
the $O'$-system is expressed in terms of that in the $O$-system  as
\beqa
\Psi_{O'}(\eta,r,\theta,\phi;\uve_c)&=&\Psi_O
(\eta,R(r,\theta),\Theta(r,\theta),\phi)
\nonumber\\
&=&\Psi_O(\eta,r,\theta,\phi;\uve_c)+\p_r\Psi_O(\eta,r,\theta,\phi;\uve_c)
(R(r,\theta)-r)\nn\\
& &+\p_\theta
\Psi_O(\eta,r,\theta,\phi;\uve_c)
(\Theta(r,\theta)-\theta)+O(d^2).\lab{ppot}
\eeqa

\subsection{Quadrupole Moment at a Cluster}
In subsection A, we have discussed the relation between the
polarization and temperature anisotropies.
Here we calculate the  quadrupole anisotropy
of CMB at a cluster $O'$ which is observed as the Stokes parameter
$Q-iU$ in the direction from the Earth $O$.
 The quadrupole mode ${\cal T}_{22}$ seen at the cluster $O'$
is written as 
\beq
(Q-iU)\propto{\cal X}(\uve_c,d)\equiv\int {\cal{T}}
(\eta_0-d,d\uve_c,\veg)Y_{22}^*(\veg;\uve_c)d\Omega_\veg ,
\label{pol}
\eeq
where we have explicitly denoted  the  orientation of the polar axis
$\uve_c$ to show our specific choice of the coordinate assigned  for
each cluster $O'$. 

The large angle (small $l$) temperature anisotropies
are dominated by the Sachs-Wolfe effect and the Integrated Sachs-Wolfe
effect. The Sachs-Wolfe effect is written in terms of the potential field
$\Psi$ in the $O'$-coordinates as
\beq
{\cal T}_{SW}(\eta_0-d,d\uve_c,\veg)=\frac13 \Psi_{O'}
(\eta_{dec},(\eta_0-d-\eta_{dec})\veg).\lab{tsw}
\eeq
With equations (\ref{ppot}), (\ref{pol}) and (\ref{tsw}) we can evaluate
the quadrupole mode ${\cal X}_{SW}$ due to the SW effect. 

 To evaluate the integral (\ref{pol}) we expand the
linear potential field $\Psi$ in the $O$-system by the spherical
harmonics as
\beq
\Psi_O(\eta,r,\theta,\phi;\uve_c)=\sum_{lm} F(\eta)
\Psi_{lm}(r;\uve_c)Y_{lm}(\theta,\phi;\uve_c).
\eeq
Here the function $F(\eta)$ represents the time dependence of the linear
potential fluctuation and is proportional to $D/a$ ($D$ is the  linear
growth rate of density 
contrast). In the case of the Einstein de-Sitter universe or at an early
matter-dominated stage of general models, we have $D/a=const$
\cite{lss}. Thus we fix 
$D/a=1$ at $\eta=\eta_{dec}$ and normalize the time dependence by
$F(\eta_{dec})=1$.

After some algebra we obtain the following result: 
\beqa
{\cal X}_{SW}(\uve_c,d)&=&\frac{\Psi_{22}(\eta_0;\uve_c)}3
\nonumber\\
&&+\frac1{3}d\lkk
-\Psi'_{22}(\eta_0;\uve_c)+\frac1{\sqrt{7}} 
\Psi'_{32}(\eta_0;\uve_c)+\frac4{\sqrt{7} B}
\Psi_{32}(\eta_0;\uve_c)\cot\frac{\eta_0}{B}\rkk+O(d^2),\lab{xsw}
\eeqa
where the prime denotes the radial derivative and we have put
$\eta_0-\eta_{dec}=\eta_0$ (since $\eta_{dec}\ll\eta_0$),  and 
have used formulas for the spherical harmonics  like
\beq
\cos\theta Y_{22}(\theta,\phi)=\frac1{\sqrt{7}}Y_{32}(\theta,\phi).
\eeq

Similarly the quadrupole mode induced by the Integrated-Sachs-Wolfe
effect is written as
\beqa
{\cal X}_{ISW}(\uve_c,d)&=&2\lop\Psi_{22}'(\eta;\uve_c)\rop+2d\Big
( -\lop\Psi_{22}'(\eta;\uve_c)\rop+\frac1{\sqrt 7}
 \lop \Psi_{32}'(\eta;\uve_c)\rop\nn\\
& &+
\frac4{{\sqrt 7}B} 
\lop{\Psi_{32}(\eta;\uve_c)}\cot\frac{\eta}{B}\rop\Big)+O(d^2), \lab{xisw}
\eeqa
where we have defined an integral operator $\lop\cdot\rop$ as
\beq
\lop y(\eta)\rop \equiv \int_0^{\eta_0} F'(\eta_0-\eta) y(\eta)d\eta.
\eeq
The $d^0$-th order term is essentially the same as the result obtained
 in \cite{ss}.
The formulas (\ref{xsw}) and (\ref{xisw}) are easily extended to general
background geometry. We obtain the formulas for the flat universe
 in the limit
$B\to \infty$ and for the open universe
 by replacement $\cot[x] \to \coth[x]$.

In the Einstein de-Sitter Universe the ISW effect vanishes
($F'=0$) and the quadrupole moment is determined solely by the
three-dimensional information of the potential field $\Psi$ at 
the decoupling.  In this case, we have the following result up to
the second-order of $d$:
\beqa
3{\cal X}(\uve_c,d)&=& \Psi_{22} +d\lmk -\Psi'_{22}  
+\frac1{\sqrt 7}\Psi'_{32} +\frac4{\sqrt 7} \frac{\Psi_{32}}{\eta_0} \rmk 
 +d^2\Big(
\frac47\Psi''_{22} +\frac1{7{\eta_0}}\Psi'_{22} 
-\frac3{7{\eta_0^2}}\Psi_{22}\\\nn
& &-\frac1{\sqrt 7}\Psi''_{32} -\frac4{{\sqrt
7}{\eta_0}}\Psi'_{32}+ \frac4{\sqrt
7 {\eta_0^2}}\Psi_{32}
+\frac1{7\sqrt3}{\Psi''_{42}}+\frac{3\sqrt 3}{7{\eta_0}}\Psi'_{42}
+\frac{5\sqrt 3}{7{\eta_0^2}}\Psi_{42} \Big)+O(d^3),
\lab{eds}
\eeqa
where we have denoted $\Psi_{lm}(\eta_0;\uve_c)$ by $\Psi_{lm}$
for simplicity.

\subsection{All Sky Map and Reconstruction of the Linear Potential Field}
The result given in the previous subsection depends on the specific
choice of the coordinate system defined for each cluster. We have set
the direction of $\theta=0$ toward the direction of the cluster
$\uve_c\propto {\overrightarrow {OO'}}$. Here in order to consider an all sky map
of the quadrupole moments seen at clusters, we express the coefficient
$\Psi_{lm}(r;\uve_c)$ in terms of the harmonic coefficients for
a spherical coordinate system fixed at the Earth.
We denote the direction of a cluster
$\uve_c$ in this fixed coordinate as 
(see figure 3)\footnote{We choose
the direction of $\phi=0$ in the $O'$-system parallel to
 ${\vect E}_\theta  $ in figure 3.}
\beq
\uve_c=(\sin\theta_c\cos\phi_c,\sin\theta_c\sin\phi_c,\cos\theta_c),
\eeq
and the linear potential field as
\beq
\Psi(\eta,r,\theta,\phi)=\sum_{lm}
F(\eta)\Psi_{lm}(r)Y_{lm}(\theta,\phi) .
\eeq

Relation between the two coefficients
$\Psi_{lm}(r;\uve_c)$ and  $\Psi_{lm}(r)$  is
given by the two successive rotations of the system by
 $\veR_z(-\phi_c)$ and $\veR_y(-\theta_c )$, where $\veR_a(\alpha)$
is the rotation operator around the $a$-axis with angle $\alpha$.
\beqa
\Psi_{l2}(r;\uve_c)&=&\lla l2 |\veR_y(-\theta_c )
\veR_z(-\phi_c)|\Psi(r,\theta,\phi)\rra\\
&=&\sum_{l'm'}\lla l2| \veR_y(-\theta_c )
\veR_z(-\phi_c)|l'm'\rra \lla l'm' |\Psi(r,\theta,\phi)\rra\\
&=&\sum_{m'} \sqrt{\frac{4\pi}{2l+1}}
\Psi_{lm'}(r){}_{-2}Y_{lm'}(\theta_c,\phi_c).\lab{spin}
\eeqa
Here we have followed the notation of Sakurai \cite{sakurai}. Namely,
the ket-vector $|f\rangle$ represents an angular function
$f(\theta,\phi)$ and bra-vector $\langle f |$ its complex conjugate. The
inner-product denotes the angular integrals,
\beq
\lla g|f\rra\equiv \int_0^{2\pi}d\phi\int_0^\pi d\theta \sin\theta
g^*(\theta,\phi) f(\theta,\phi).
\eeq
The function  ${}_{-s}Y_{lm}(\theta,\phi)$ in the equation (\ref{spin})
is the spin-weighted spherical harmonic \cite{newman} and 
is given in terms of the rotation matrix of the scalar spherical
harmonic $|lm\rangle \equiv Y_{lm}$ as
\beq
{}_{-s}Y_{lm}(\theta,\phi)\equiv 
 \sqrt{\frac{2l+1}{4\pi}} \lla l s|\veR_y(-\theta)
\veR_z(-\phi)|lm\rra.
\eeq
Thus the all sky (three-dimensional) map of 
${\cal X}={\cal X}_{SW}+{\cal X}_{ISW}$ is 
expressed in terms of the coefficients
$\Psi_{lm }$ defined on a fixed coordinate system  as
\beqa
{\cal X}_{SW}(\uve_c,d)&=&\sqrt{\frac{4\pi}{45}}{}_{-2}Y_{2m}(\theta_c,\phi_c)
\Psi_{2m }(\eta_0)+d\Big[
-\sqrt{\frac{4\pi}{45}}{}_{-2}Y_{2m}(\theta_c,\phi_c)
\Psi_{2m }'(\eta_0)\nn\\
& &+\frac{\sqrt {4 \pi}}{21} \lmk\Psi_{3m}'(\eta_0)
+\frac4B\Psi_{3m }(\eta_0) \cot\frac{\eta_0}B \rmk
{}_{-2}Y_{3m}(\theta_c,\phi_c) 
\Big]+O(d^2),\lab{psw}
\eeqa 
and
\beqa
{\cal
X}_{ISW}(\uve_c,d)&=&2\sqrt{\frac{4\pi}{5}}{}_{-2}Y_{2m}(\theta_c,\phi_c) 
\lop\Psi_{2m }(\eta)\rop+2d\Big[
-\sqrt{\frac{4\pi}{5}}{}_{-2}Y_{2m}(\theta_c,\phi_c)
\lop\Psi_{2m }'(\eta)\rop\nn\\
& &+\frac{\sqrt {4 \pi}}{7}
\lmk\lop\Psi_{3m}'(\eta)\rop+\frac4B\lop\Psi_{3m }(\eta)
\cot\frac{\eta}B \rop \rmk
{}_{-2}Y_{3m}(\theta_c,\phi_c) 
\Big]+O(d^2).\lab{pisw}
\eeqa 

Assuming that we have the all sky map of $\cal X$,
we can extract out information of $\Psi_{lm}$ for each $(l,m)$-mode
by using the orthonormal relation for the spin-weighted harmonics,
\beq
\int_0^{2\pi}d\phi\int_0^\pi d\theta 
\sin\theta\, {}_{-s}Y_{lm}^*(\theta,\phi)\,{}_{-s}Y_{l'm'}(\theta,\phi)
=\delta_{l'l}\delta_{m'm}\,.
\eeq
{}From equations (\ref{psw}) and (\ref{pisw}), we see that by
observing the $d$-dependence (or equivalently the linear redshift 
dependence) of the polarization map, we obtain information of the linear
potential fluctuations in the combination,
\beq
-\sqrt{\frac{4\pi}{45}}
\Psi_{2m }'(\eta_0)-2\sqrt{\frac{4\pi}{5}}\lop
\Psi_{2m }'(\eta_0)\rop,
\label{ell2}\eeq
for $l=2$ modes
and
\beq
 \frac{\sqrt {4 \pi}}{21} \lmk\Psi_{3m
}'(\eta_0)+4\Psi_{3m }(\eta_0) \cot\frac{\eta_0}B \rmk
+\frac{\sqrt {4 \pi}}{7} \lmk\lop\Psi_{3m
}'(\eta)\rop+\frac4B\lop\Psi_{3m }(\eta) \cot\frac{\eta}B \rop \rmk,
\label{ell3}\eeq
for $l=3$ modes.

In the case of the Einstein de-Sitter universe, the multipole components
 $\Psi_{lm}(\eta_0)$ for small $l$ at the last scattering surface
 $r=\eta_0$ can be obtained directly from the temperature anisotropies
 $\cal T$. If we have the polarization map we can further obtain the 
following information: From the $O(d)$ map and the temperature
anisotropy $\cal T$, we obtain
\beq
\Psi_{2m}'(\eta_0),~~\Psi_{3m}'(\eta_0),
\eeq
and from the  polarization map up to $O(d^2)$ order and the temparature
anisotropy $\cal T$,
\beq
 \Psi_{2m}''(\eta_0), ~~\Psi_{3m}''(\eta_0),~~
9\frac{\Psi_{4m}'(\eta_0)}{\eta_0}+\Psi_{4m}''(\eta_0). 
\eeq
Note that we cannot separate $\Psi_{4m}'(\eta_0)$ from
$\Psi_{4m}''(\eta_0)$.
Using information of the polarization up to the $d^n$ (or $z^n$)-order we
 would know the derivative coefficients
 $\p_{\eta_0}^i\Psi_{2m}(\eta_0) $  and 
 $\p_{\eta_0}^i\Psi_{3m}(\eta_0) $ separately for  $i\le n$ but not
 for $l\ge 4$ modes.

So far we have inquired a basic problem of reconstruction,
considering an idealistic and simplified situation. Here we
have to mention two points that would be important to apply
 our reconstruction method  to real observational data. 
The first point is that the polarization signal $Q-iU$ of
 a cluster is obtained by a combination of the temperature
 anisotropy ${\cal T}_{22}$ at the cluster and its optical
 depth $\tau$, as shown in equation (12). We
have implicitly assumed that we can  separate them. 
This could be achieved, for example, by using X-ray and thermal
 Sunyaev-Zeldovich data.
The second point is that distribution of clusters is not
homogeneous on our light cone (both in redshift and angular
 position).   To extract out information of the potential
 field from inhomogeneous sample of clusters it would be
 necessary to develop a workable statistical method.

\section{Examples}
As shown in the previous section the  polarization signal contains both
 the SW  and ISW effects. The former  reflects local quantity
at the last scattering surface and more interesting from the point of
reconstructing the linear potential field $\Psi$.  

In this section we calculate the
magnitude of these two effects for concrete models. We investigate flat
cold-dark-matter (CDM) models with cosmological constant $\lambda_0$
$(\Omega_0+\lambda_0=1)$. We use the primordially
 Harrison-Zeldovich spectrum with the  CDM
transfer function  given in Bardeen et al. \cite{bardeen}. The Hubble
parameter
is fixed at $h=0.7$ ($h$ is the Hubble parameter in units of
100km/sec/Mpc). The shape parameter $\Gamma$ of the CDM transfer function is
set  at
$\Gamma=h\Omega_0$.

We first calculate the rms value for the $d^0$-th signal of ${\cal X}$
(eqs.(\ref{psw}) and (\ref{pisw})),
\beq
-\sqrt{\frac{4\pi}{45}}
\Psi_{2m }(\eta_0)-2\sqrt{\frac{4\pi}{5}}\lop
\Psi_{2m }(\eta_0)\rop,
\eeq
and the same quantities for the $d^1$-st order correction for $l=2$ mode
both for the SW and ISW effects (eq.(\ref{ell2}))
\beq
-\sqrt{\frac{4\pi}{45}}
\Psi_{2m }'(\eta_0),~~~-2\sqrt{\frac{4\pi}{5}}\lop
\Psi_{2m }'(\eta_0)\rop.
\eeq
For $l=3$ mode at the $d^1$-st order the SW effect is constituted by two
terms (eq. (\ref{ell3})) and we treat them  separately. We evaluate the
rms 
values of the three quantities,
\beq
 \frac{\sqrt {4 \pi}}{21} \Psi_{3m}'(\eta_0),
~~~4 \frac{\sqrt {4\pi}}{21\eta_0}\Psi_{3m }(\eta_0),
~~~\frac{\sqrt {4 \pi}}{7} \lmk\lop\Psi_{3m}'(\eta)\rop
+4\lop\frac{\Psi_{3m }(\eta)}{\eta}  \rop \rmk.
\eeq
In figure 4 we plot the ratio of the $d^1$-st signal at $z=0.5$ to
the $d^0$-th signal.
First, note that the  magnitude of the $d^1$-st correction is significant
even at $z=0.5$. For $l=2$ mode the SW effect is larger than the ISW
effect for $\Omega_0\gsim 0.3$
 and the $d^1$-st signal becomes comparable to the $d^0$-th signal for
 $\Omega_0\simeq 1.0$. 
For $l=3$ modes, the $\Psi_{3m}$ contribution to the SW effect is
larger than the $\Psi'_{3m}$ contribution. Here the ISW effect is
stronger than the case 
of $l=2$ and the SW effect becomes dominant only for
$\Omega_0\gsim0.8$.

\section{summary}

The polarization signal of CMB associated with a distant cluster reflects
temperature anisotropies observed at the cluster.  Using this effect we
can, in principle, obtain three-dimensional information of potential
fluctuations  around the last scattering surface (LSS).
In this paper we have calculated the  three-dimensional map of the
polarization signal of clusters induced by the temperature anisotropies
in homogeneous and isotropic background universe. We have considered
adiabatic scalar perturbations and
included both the Sachs-Wolfe effect and the Integrated-Sachs-Wolfe
effect.  Our formulation is valid for general background geometries.
 The radial part of the three-dimensional map is expressed
perturbatively with respect to the distances (or equivalently redshifts)
of clusters, and angular dependence is written in terms of the
spin-weighted harmonics. Using the orthonormality of the harmonics, the
independent information that can be extracted  out from the  map
 is clarified explicitly.

In the case of the Einstein de-Sitter universe, the
Integrated-Sachs-Wolfe
effect vanishes and the temperature anisotropies solely probe the local
quantities around the LSS.  We showed that using the polarization signal
up to the $z^n$-th order, the derivative coefficients
 $\p^i_{\eta_0}\Psi_{lm}$ would be separately 
obtained for $l=2$ and 3 modes. But we cannot separate them  for
modes with $l\geq4$.

For general background cosmological models the polarization signal
also contains the Integrated-Sachs-Wolfe effect. In \S III we have
examined the first ($z^1$-th) order 
 polarization signal for typical flat CDM models
with cosmological constant. We found that the Sachs-Wolfe effect
dominates the signal of $l=2$ modes  for density parameter
 $\Omega_0\gsim 0.3$, but it dominates the $l=3$ signals only for
$\Omega_0\gsim 0.8$.
 The Integrated-Sachs-Wolfe effect is nonlocal effect and
troublesome from the point of view to reconstruct the primordial
potential fluctuations.
But it mainly comes from time variation of the potential field at 
relatively recent epoch. Hence we might be able to separate its effect
by probing the large-scale matter distribution with other observational
tools ({\it e.g.} Ref.\cite{isw}). 

\acknowledgments
NS  would like to thank  N. Sugiyama and G.-C. Liu 
for useful discussions.
This work has been  supported by the Japanese Grant
in Aid for Science Research Fund
of the Ministry of Education, Science,
Sports and Culture
Nos. 03161(NS) and 12640269(MS).

\newpage
\begin{figure}[h]
 \begin{center}
 \epsfxsize=10cm
 \begin{minipage}{\epsfxsize} \epsffile{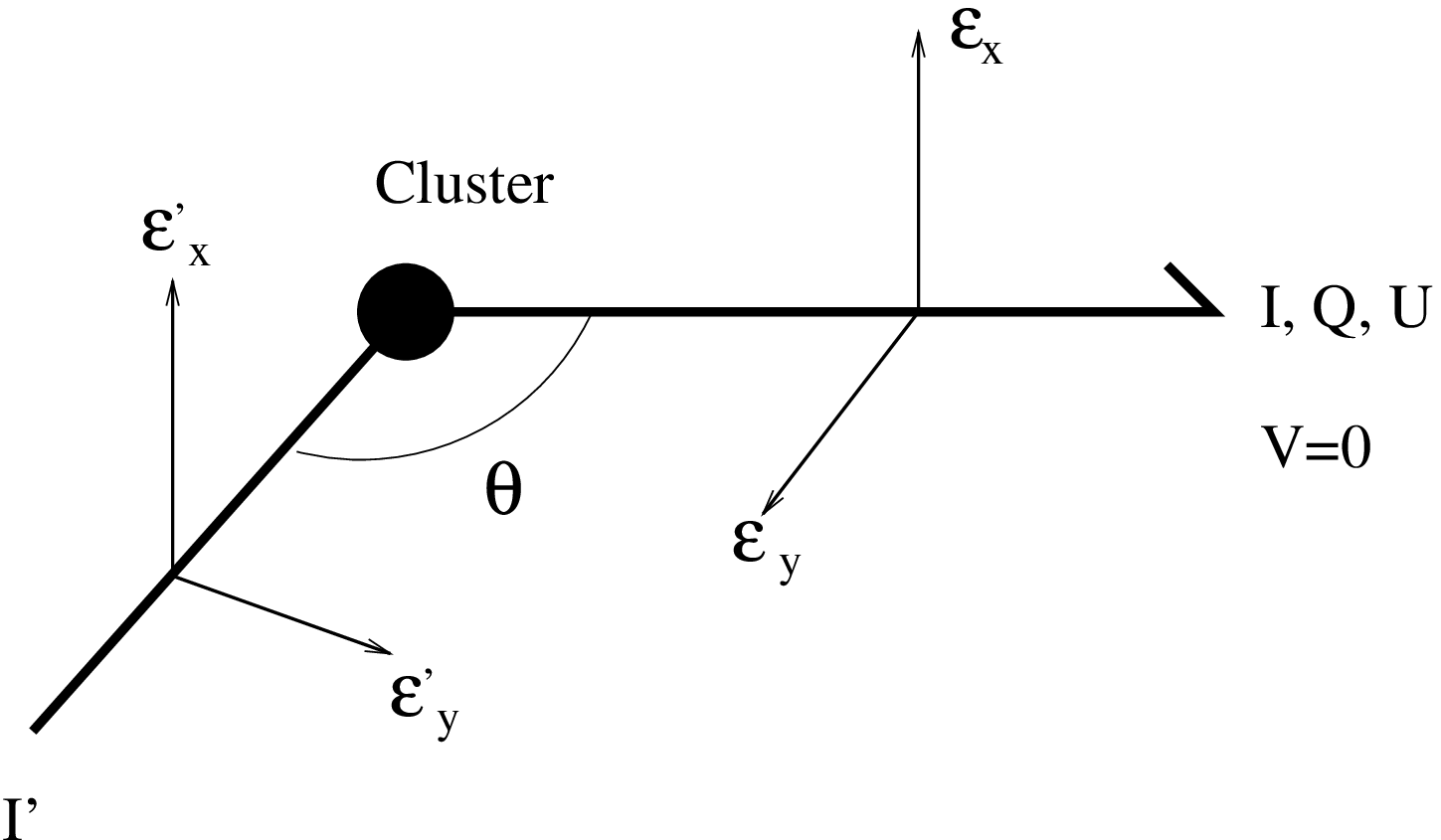} \end{minipage}
 \end{center}
\caption[]{ The Thomson scattering of the CMB
 photon with ionized gas of a cluster. Initially unpolarized radiation
  becomes polarized by the quadrupole anisotropy of the incident wave
 $I'$.}
\end{figure}

\begin{figure}[h]
 \begin{center}
 \epsfxsize=10cm
 \begin{minipage}{\epsfxsize} \epsffile{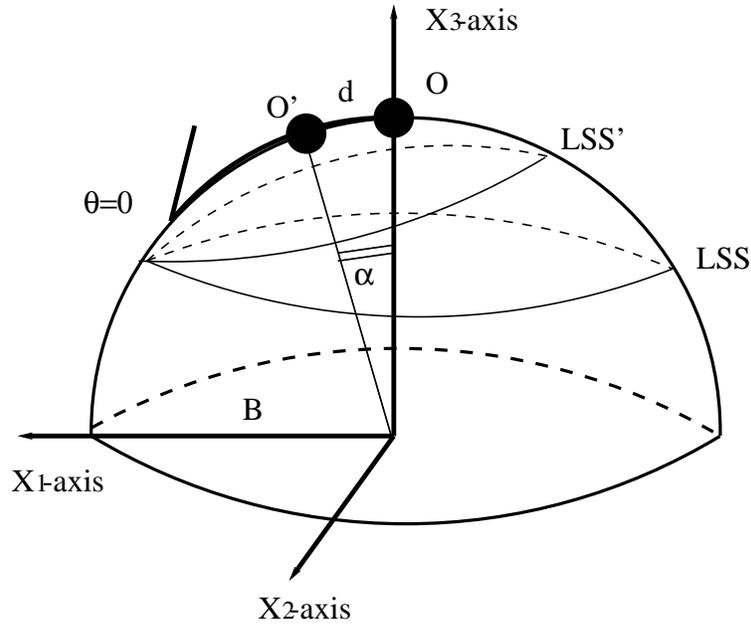} \end{minipage}
 \end{center}
\caption[]{ Correspondence of the two coordinate systems $(R,\Theta)$ and
 $(r,\theta)$ centered at 
 different places $O$ and $O'$, respectively.  We consider a closed
 universe with 
 curvature radius  $B$. Two systems coincides with each other by
 rotation of  angle $\alpha\equiv d/B$ around the $X_2$-axis.}
\end{figure}

\begin{figure}[h]
 \begin{center}
 \epsfxsize=10cm
 \begin{minipage}{\epsfxsize} \epsffile{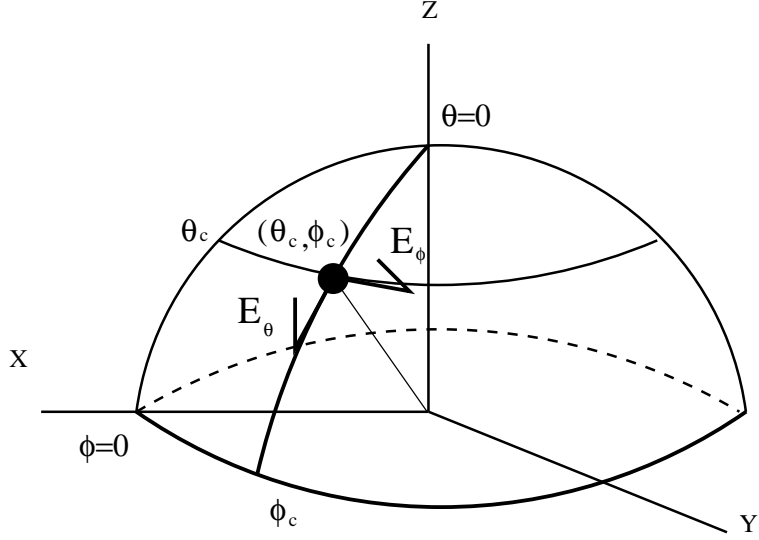} \end{minipage}
 \end{center}
\caption[]{ Relation between the  coordinate system specific to each
 cluster and the Earth fixed coordinate system.}
\end{figure}

\begin{figure}[h]
 \begin{center}
 \epsfxsize=10cm
 \begin{minipage}{\epsfxsize} \epsffile{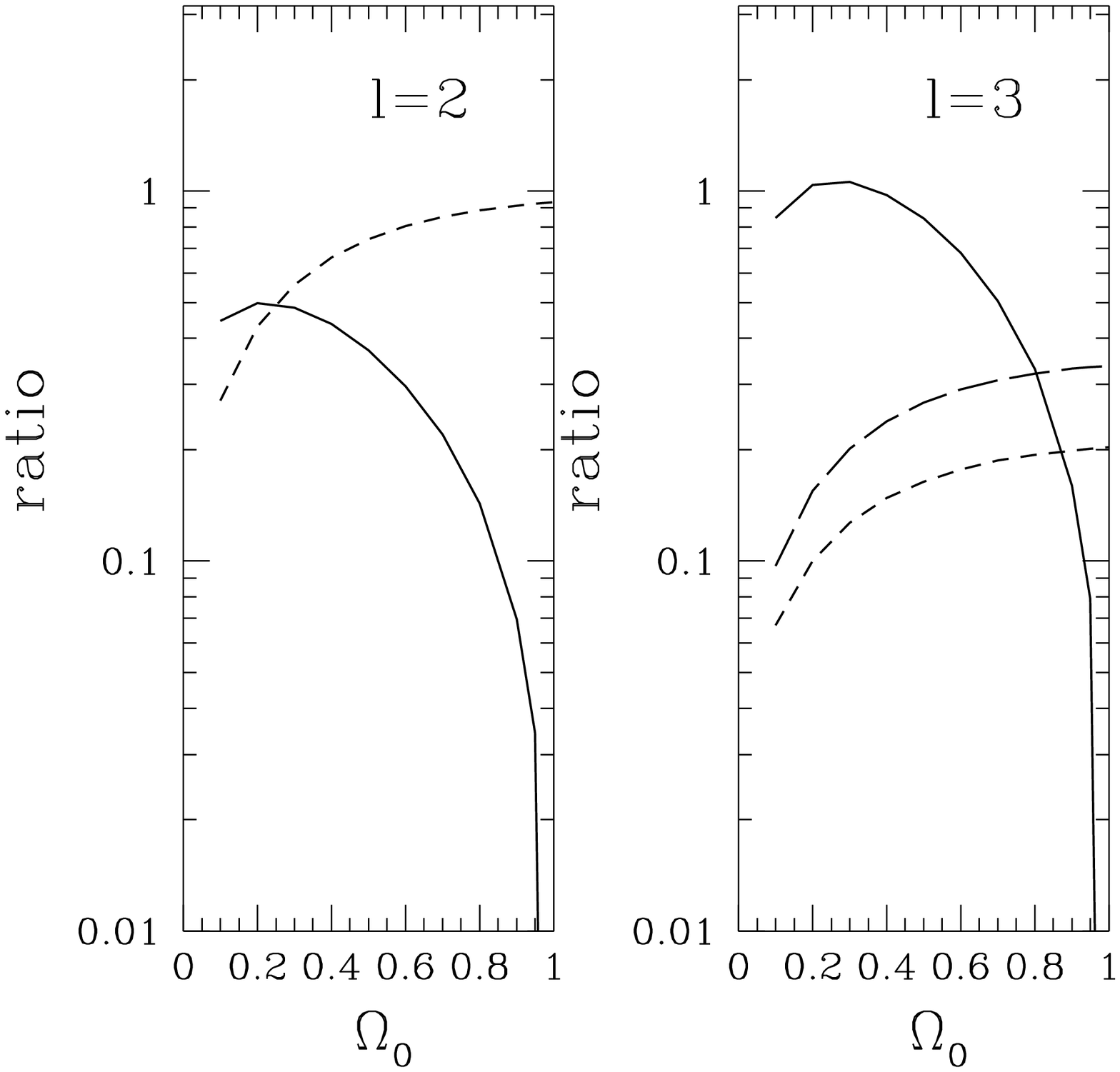} \end{minipage}
 \end{center}
\caption[]{Ratio of the $d^1$-st order polarization signal and the
 $d^0$-th signal for a cluster  at $z=0.5$. We consider flat CDM models
 with cosmological constant.  The dashed lines represent the contribution
 of the SW effect and the solid of  the ISW effect.
For $l=3$ modes, there are  two SW terms.  The long-dashed line
 represents the term proportional to  $ \Psi$ and the 
 short-dashed line  the term proportional to $ \Psi'$. }
\end{figure}


\begin{references}

\bibitem{zelsun}Ya. Zeldovich  and R. A.  Sunyaev, Sov. Astron. Lett.\
 {\bf 6}, 285 (1980).

\bibitem{koso}A. Kosowsky, Ann.Phys. (N.Y.)  \ {\bf
246}, 49 (1996).

\bibitem{ss}S. Y. Sazonov  and R. A.  Sunyaev, Mon. Not, R. Astron. Soc.
 {\bf 310}, 765 (1999).

\bibitem{kogu}A. Kogut et al., Astrophys. J. Lett.
 {\bf 464}, 5 (1996).

\bibitem{kl}M. Kamionkowski  and A. Loeb, Phys. Rev, D,
 {\bf 56}, 4511 (1997).

\bibitem{zal1}M. Zaldarriaga and U. Seljak, Phys. Rev, D,
 {\bf 55}, 1830 (1997).

\bibitem{kami1}M. Kamionkowski A. Kosowsky and A. Stebbins, Phys. Rev, D,
 {\bf 55}, 7368 (1997).

\bibitem{hu1}W. Hu  and M. White, Phys. Rev, D,
 {\bf 56}, 596 (1997).

\bibitem{chandra}S.  Chandrasekhar, {\it Radiative Transfer}
 (Dover, New York, 1960).

\bibitem{radi}G. B. Rybicki and A. P. Lightman, {\it Radiative Process
 in Astrophysics}
 (Wiley, New York, 1960).

\bibitem{zal2}M. Zaldarriaga, Phys. Rev, D,
 {\bf 55}, 1822 (1997).

\bibitem{kodsas} H. Kodama and M. Sasaki, Int. J. Mod. Phys. A,
 {\bf 1}, 265 (1986).

\bibitem{sw}R. K. Sachs  and A. M. Wolfe, Astrophys. J.
 {\bf 147}, 73 (1967).

\bibitem{gouda}N. Gouda, N. Sugiyama  and M. Sasaki, Prog. Theor. Phys.\
 {\bf 85}, 1023 (1991).

\bibitem{lss}P. J. E.  Peebles, {\it The Large Scale Structure in the
 Universe}
 (Princeton University Press, Princeton, 1980).

\bibitem{sakurai}J. J.  Sakurai, {\it Modern Quantum Mechanics}
 (Addison-Wesley, New York, 1985).

\bibitem{newman}E. Newman and R. Penrose, 
J. Math.\ Phys. (N.Y.)\ {\bf 7,} 863 (1966), J. N. Goldberg et al., {\it
 ibid},\ {\bf 8,} 2155 (1967)

\bibitem{bardeen}J. M. Bardeen J. R. Bond N. Kaiser  and A. Szalay,
  Astrophys. J. {\bf 304}, 15 (1986).

\bibitem{isw}R. G. Crittenden and N. G. Turok, Phys. Rev, Lett,
 {\bf 76}, 575 (1996) A. Kinkhabwala and  M. Kamionkowski, {\it
 ibid},
 {\bf 82}, 4172 (1999).


\end{references}
\end{document}